\documentclass[aip,preprint,floatfix]{revtex4-1}
\usepackage{graphicx}
\usepackage{amsmath}
\usepackage{hyperref}

\usepackage{siunitx}
\newcommand{\NbOx}{NbO$_\mathrm{x}$}
\newcommand{\NbOt}{NbO$_\mathrm{2}$}
\newcommand{\NbtOf}{Nb$_\mathrm{2}$O$_\mathrm{5}$}
\newcommand{\degC}{$^\circ$C}

\newcommand{\vth}{\ensuremath{V_\mathrm{th}}}%
\newcommand{\vho}{\ensuremath{V_\mathrm{h}}}%

\newcommand{\pth}{\ensuremath{P_\mathrm{th}}}%

\begin{document}


\title{ Enhancement in neuromorphic NbO$_\mathrm{2}$ memristive device switching at cryogenic temperatures} 



\author{Ted Mburu}
\affiliation{Department of Physics and Astronomy, Ithaca College, Ithaca NY}

\author{Zachary R.\ Robinson}
\affiliation{Department of Physics, SUNY Brockport, Brockport, NY}

\author{Karsten Beckmann}
\affiliation{College of Nanotechnology, Science, \& Engineering, University at Albany, Albany, NY}
\affiliation{NY CREATES, Albany NY}

\author{Uday Lamba}
\affiliation{Department of Physics and Astronomy, Ithaca College, Ithaca NY}

\author{Alex Powell}
\affiliation{Department of Physics and Astronomy, Ithaca College, Ithaca NY}

\author{Nathaniel Cady}
\affiliation{Colleges of Nanotechnology, Science, \& Engineering, University at Albany, Albany, NY}

\author{M.\ C.\ Sullivan}
\email[]{mcsullivan@ithaca.edu}
\affiliation{Department of Physics and Astronomy, Ithaca College, Ithaca NY}


\date{\today}

\begin{abstract}
The electrical properties and performance characteristics of niobium dioxide (\NbOt)-based memristive devices are examined at cryogenic temperatures.  Sub-stoichiometric  \NbtOf{} was deposited via magnetron sputtering and patterned in microscale (2$\times$2 - 15$\times$15 $\mu$m$^2$) cross-bar Au/Ru/\NbOx/Pt devices and electroformed at 3-5 V to make \NbOt\ filaments.  At cryogenic temperatures, the threshold voltage (\vth) increased by more than a factor of 3.  The hold voltage (\vho) was significantly lower than the threshold voltage for fast voltage sweeps (200 ms per measurement).  If the sample is allowed to cool between voltage measurements, the hold voltage increases, but never reaches the threshold voltage, indicating the presence of non-volatile \NbtOf\ in the filament.  The devices have an activation energy of $E_a \approx 1.4$ eV, lower than other \NbOt\ devices reported.  Our works shows that even nominally ``bad" memristive devices can be improved by reducing the leakage current and increases the sample resistance at cryogenic temperatures.

\end{abstract}

\pacs{}

\maketitle 


\section{Introduction}










As the demand for computational power increases, the energy efficiency and scaling limitations of conventional CMOS-based circuits have become increasingly apparent.  The rapidly developing field of neuromorphic computing seeks to overcome these limitations by emulating the functions of the human brain.\cite{Islam2023}  The goal of neuromorphic computing is to revolutionize memory storage and computational paradigms by offering non-volatile storage and analog computing capabilities within the same architecture.  In order to realize the goals of neuromorphic circuits, materials must be found that can mimic both the synaptic and neuronal behaviors in nerve cells.\cite{Indiveri_2013, Jo_2010_NanoLet, Yang_2013_NatureNano, wright2013,hasler2013finding,pickett2013scalable}

Crystalline niobium dioxide has emerged as a material of significant interest in neuromorphic circuits, for example, as a scalable neuristor,\cite{pickett2013scalable} as selectors in non-volatile RAM,\cite{Gibson_APL_2016} or as oscillator neuron devices.\cite{Kwon_2024}  \NbOt\ has a structural phase transition at approximately $800$\degC\ that is marked by a sharp insulator-metal transition (IMT), dramatically changing the resistance of the material.\cite{1969582,JANNINCK19661183, wahila_evidence_2019, paez2021structural}  Reliable threshold switching about the IMT enables the precise control of the device's resistance state, a prerequisite for the reliable operation of neuromorphic circuits.\cite{Liu2014}

The preferred oxidation state of niobium oxide is \NbtOf, which makes deposition of crystalline \NbOt\ challenging.  While many studies successfully deposit \NbOt\ films  (sometimes crystalline\cite{Park_APL2016, Park_SR2017} though more often amorphous or poly-crystalline\cite{Kumar2017_nature, Joshi_JAP_2018, Stoever2020}),  most researchers deposit \NbtOf\ \cite{Li_2018_Nano, Nath_Nanotech_2020, Nath_ACSAMI_2021} or a mixture of NbO, \NbtOf, and \NbOt\ (usually referred to as \NbOx) \cite{Gibson_APL_2016, Kozen_ACSAMI_2020, Chen_IEEED_2021, Chen_Ceramics_2021, Sullivan2022}, or intentionally layer \NbOt\ and \NbtOf \cite{Liu2014, Nandi_2015_JAPD, Nandi_ACSAMI_2020, Lee_JVST_2021} or layer NbO and \NbOx \cite{Park_RSC_2022}.  In some reports, the exact composition of the \NbOx\ films is unclear.\cite{Chen_2018_IEEETrans, Wang_2020_APL, Kwon_2024}

If the initially deposited film is not \NbOt, researchers must create \NbOt\ from the as-deposited film.  The most straightforward method is post-deposition annealing,\cite{Kumar2017_nature, Stoever2020, Park_RSC_2022, paez2021structural, Fridriksson2022, Sullivan2022} with a wide range of annealing temperatures, pressures, and ambient gases.  The more common method is to create an electroformed \NbOt\ filament by applying a large voltage across the film between the bottom and the top electrode.\cite{Nandi_2015_JAPD, Park_APL2016, Chen_2018_IEEETrans, Joshi_JAP_2018, Li_2018_Nano, Nath_Nanotech_2020, Nandi_ACSAMI_2020,  Chen_IEEED_2021, Nath_ACSAMI_2021, Sullivan2022, Kwon_2024}  These electroformed devices are usually a crystalline \NbOt\ filament of $r\approx 150$ nm\cite{Nandi_ACSAMI_2020} surrounded by amorphous \NbOt\ and \NbtOf\cite{Kwon_2024}.  Some researchers combine annealing and electroforming,\cite{Lee_JVST_2021} and some are unclear how they form \NbOt.\cite{Wang_2020_APL}

Finally, memristive \NbOt\ devices typically come in one of two forms.  The most common is a simple cross-bar geometry, where the \NbOx\ layer is sandwiched between two metal electrodes.  In the cross-bar geometry, the typical device sizes range from 2$\times$2 to 40$\times$40 $\mu$m$^2$. \cite{Liu2014, Nandi_2015_JAPD, Nandi_2015_APL, Park_SR2017, Chen_2018_IEEETrans, Li_2018_Nano, Nandi_ACSAMI_2020, Chen_IEEED_2021, Nath_Nanotech_2020, Wang_2020_APL, Nath_ACSAMI_2021, Park_RSC_2022, Sullivan2022, Kwon_2024, Chen2022}  These devices nearly always require the stochastic electroforming step, and the electroformed \NbOt\ filaments are nearly always much smaller than the device size.\cite{Nandi_ACSAMI_2020}  Some devices of this type rely on a percolation path through the crystalline \NbOt.\cite{Lee_JVST_2021}  Less common is a nanoscale-sized bottom electrode (20$\times$20 to 200$\times$200 nm$^2$) photolithographically patterned as a small metal via through an insulating layer, covered with a blanket niobium oxide film and a large top electrode.\cite{Gibson_APL_2016, Kumar2017_nature, Kozen_ACSAMI_2020, Lee_JVST_2021, Sullivan2022}  These films greatly constrain the device size to be about the size of an electroformed filament or smaller.

Given the wide range of \NbOx\ films and devices, it is not surprising that there is a wide range of switching behaviors in \NbOt\ memristors.  Many published reports have found threshold switching behavior with on and off-state resistances (or currents) that can change by an order of magnitude (and often more).\cite{Liu2014, Nandi_2015_JAPD, Nandi_2015_APL, Mikolajick_IEEE_2016, Park_SR2017, Chen_2018_IEEETrans, Li_2018_Nano, Nandi_ACSAMI_2020, Chen_IEEED_2021, Lee_JVST_2021, Nath_Nanotech_2020, Wang_2020_APL, Nath_ACSAMI_2021, Park_RSC_2022,Sullivan2022, Kwon_2024, Chen2022, Sullivan2022}   On occasion, researchers have reported \NbOt\ films that display non-ohmic behavior but without an abrupt switch from high to low resistance state.\cite{Chen_Ceramics_2021, Lee_JVST_2021, Sullivan2022}  Controlling the behavior of \NbOt\ to achieve repeatable and reliable memristive devices is a key goal of researchers in this field.

We have deposited films of sub-stiochiometric \NbtOf\ via physical vapor deposition and created devices in the common cross-bar geometry, with device sizes ranging from  2$\times$2 - 15$\times$15 $\mu$m$^2$.  We electroformed an \NbOt\ filament which gave poor memristive behavior at room temperature.\cite{Sullivan2022}  In this work, we examined the behavior of the \NbOt\ filaments at cryogenic temperatures.  In all cases, the threshold voltage increased and the resistive ratio between the off and on states also increased (high and low resistance states), improving the memristive device characteristics.

\section{Experiment}

All \NbOt{} films were deposited at the University at Albany College of Nanotechnology, Science \& Engineering in a Kurt J. Lesker PVD75 system via reactive magnetron sputtering using targets purchased from Kurt J.\ Lesker. The deposition occurred in an oxygen poor environment at 25\degC{} at 3\% O$_2$ flow in Ar atmosphere at 3 mTorr with a target power density of 6 W/cm$^2$. This process yields sub-stoichiometric amorphous \NbtOf{} after the deposition which can be crystallized into the metastable allotrope \NbOt{}. 

X-ray photoelectron spectroscopy (XPS) was measured with a Kratos Axis Ultra DLD, which has a monochromatic Al-k$\alpha$ X-ray source, and hemispherical analyzer. Survey spectra were measured with a pass energy of 100~eV, and high resolution spectra were measured with a pass energy of 20~eV. XPS data were analyzed with the CasaXPS software package.

Atomic force microscopy images were taken with a Park System AFM in tapping mode. The measurements were performed in air. We used the open-source  software package Gwyddion to analyze our data.

\begin{figure}
    \includegraphics[width=0.95\linewidth]{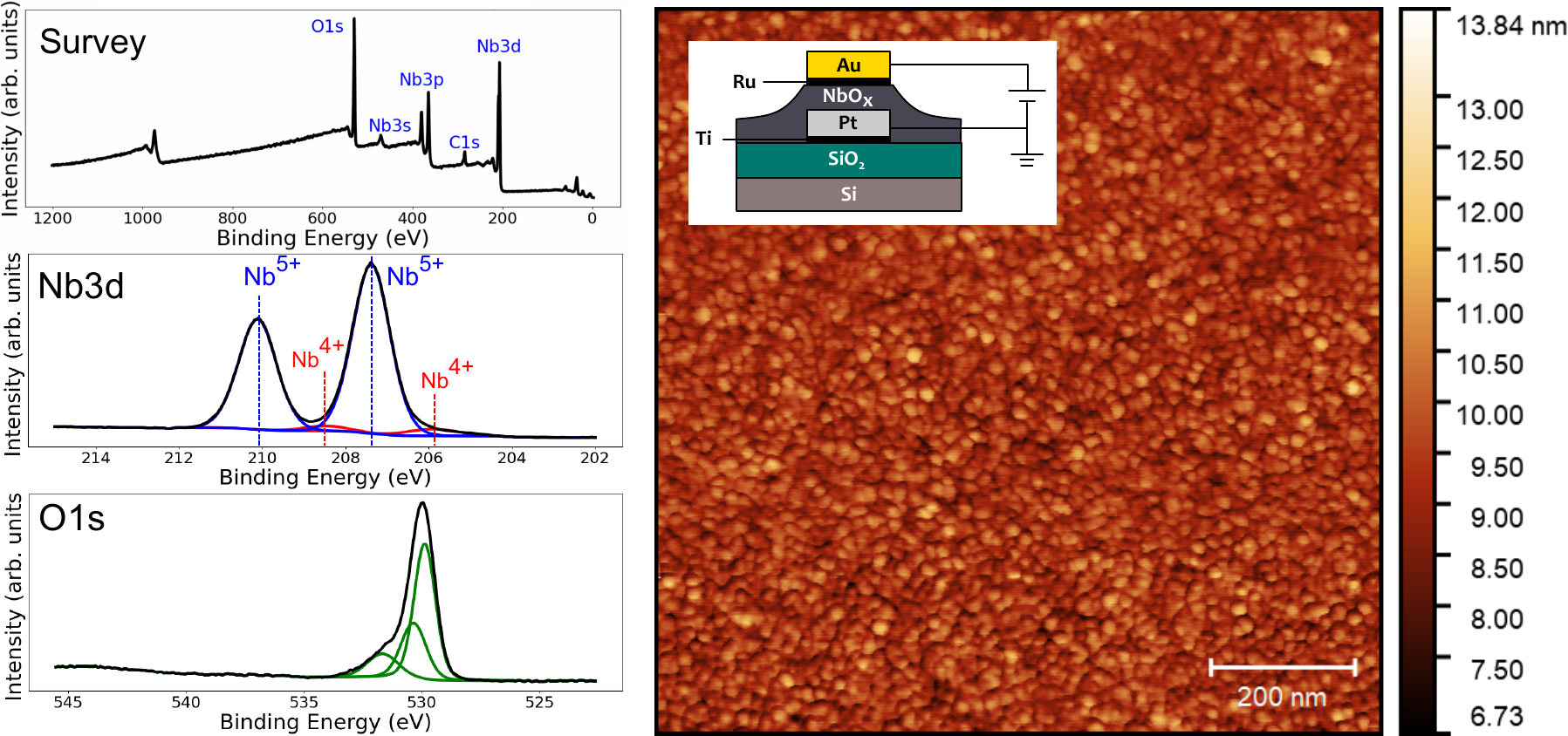}  
     \caption{Sample characterization of the \NbOt\ films.  The left shows the XPS results, indicating that the as-grown sample is mostly \NbtOf.  The right shows a representative AFM topography of a blanket as-deposited \NbtOf\ film, with RMS roughness of around 0.33~nm.  The inset on the AFM topography shows the device schematic.}
    \label{fig:DevandChar}
\end{figure}

The cross-bar device structures were fabricated at the Cornell NanoScale Science and Technology Facility with 50 nm-thick Pt bottom electrodes ranging in width from 2 $\mu$m to 15 $\mu$m on SiO$_\mathrm{2}$.  Blanket NbO$_\mathrm{x}$ deposition occurred on top of the bottom electrode, and then capped with a top electrode of a 10 nm adhesion layer of Ru followed by 50 nm of either Au or Pt.  Device patterning used standard liftoff photolithography and RIE for cleaning and to etch through the  NbO$_\mathrm{x}$ layer to reach the bottom electrode.  The schematic of the cross devices is shown as an inset in Fig.\ \ref{fig:DevandChar}.  The \NbOx\ films in the cross-bar devices were not thermally annealed after deposition. 

The electric behavior of the devices were measured using a Keithley 2400 Source Measure Unit in voltage control mode with a compliance current of 1 mA.  IV curves were initially swept from -1 to 1 V, with some sweeps reaching voltages as high as $\pm$5 V to create the electroformed \NbOt\ filaments.  The devices were cooled to low temperatures in a Janis closed-cycle cryostat.  The temperature was monitored using a Lakeshore 336 temperature controller.

\section{Results and Discussion}

\subsection{Sample growth and characterization}

The as-grown materials were characterized with XPS, XRD, and AFM.  Characterization was measured on blanket film depositions using similar deposition conditions as on the device substrates.

XPS survey scans indicate the presence of Nb, O, and adventitious C contamination.  High resolution scans for binding energies corresponding to the niobium 3d doublet can be found in Figure \ref{fig:DevandChar}, with large peaks at around 207.7~eV and 210.4~eV, which correspond to the doublet for \NbtOf{} (+5 oxidation state).  A second, vanishingly small, doublet corresponding to the +4 oxidation state of \NbOt\ occurs roughly $\sim 2.2$ eV lower than the doublet for the +5 oxidation state.  Typically, when \NbOt{} is exposed to atmospheric conditions, it will form a thin surface oxide of \NbtOf{} due to the interaction with atmospheric oxygen. \cite{wahila_evidence_2019, paez2021structural,twigg2021transmission,Robinson2022Effect} However, for \NbOt{} bulk films, the surface oxide is typically limited to the top 1.5~nm to 2.5~nm, and XPS is expected to show strong \NbOt\ doublet peaks which corresponds to the underlying \NbOt{} film \cite{Kozen_ACSAMI_2020}. The lack of a significant secondary doublet implies that the as-grown film is almost exclusively \NbtOf.

The AFM images look qualitatively similar between all samples. A representative image can be found in Figure \ref{fig:DevandChar}, with RMS roughness for the sample at 0.33~nm.  XRD measurements indicate that the as-grown films are amorphous, with crystalline \NbOt\ developing upon annealing at 600 to 900~\degC\cite{Sullivan2022}.  The films in this study were not annealed.

\subsection{Cryogenic measurements of \NbOt\ filaments}
\begin{figure}
    \includegraphics[width=.8\linewidth]{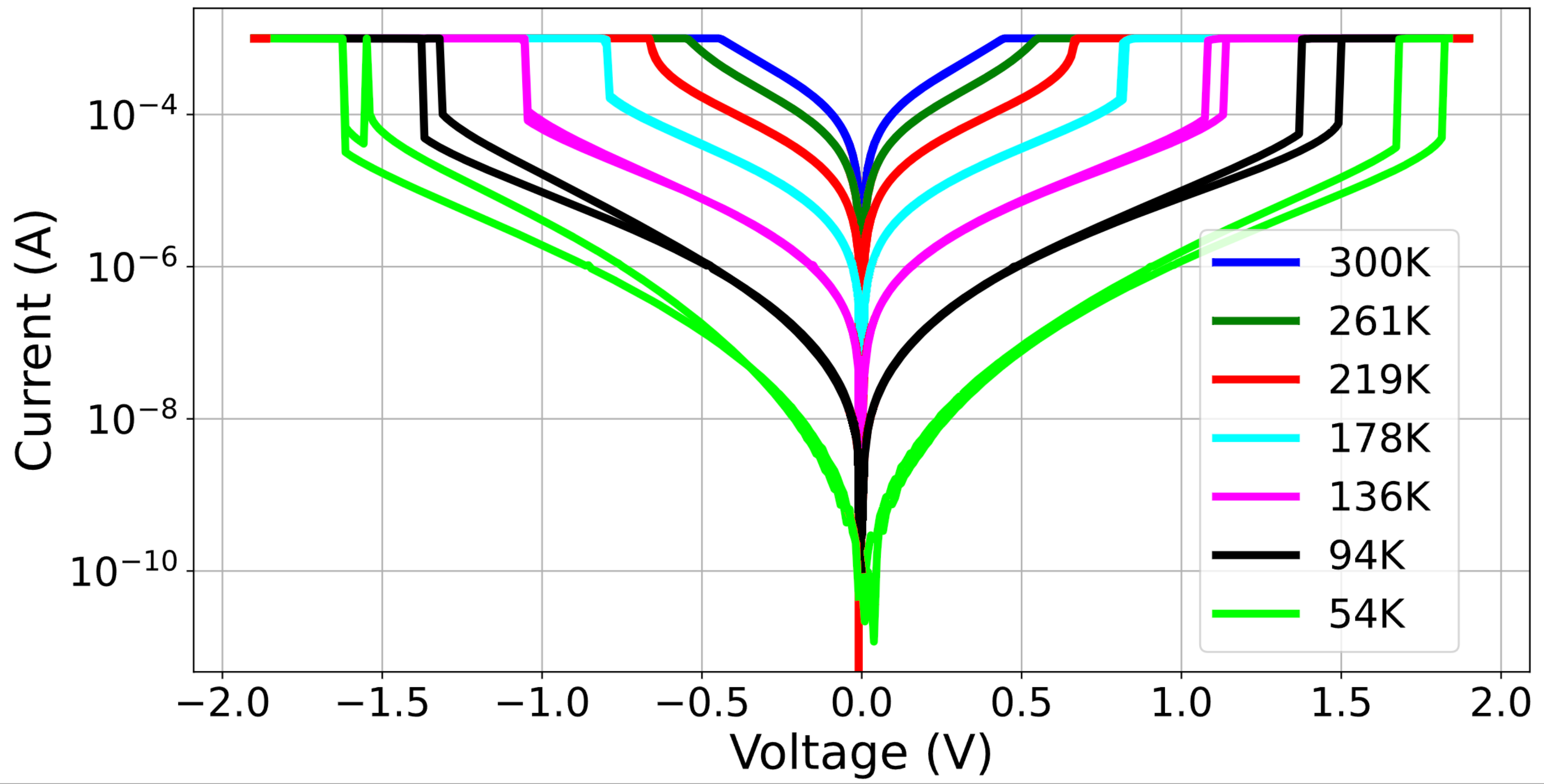} \\
    \caption{Representative I-V curves of an electroformed \NbOt\ filament measured on a $15\times 15~\mu$m$^2$ cross-bar device.  As the surrounding temperature decreases, the resistance, \vth\, and $R{_\textrm{on/off}}$ increases.  The input power was turned off after each measurement, allowing the sample to cool.}
    \label{fig:IVs}
\end{figure}

We present our results on cryogenic measurements on a typical electroformed \NbOt\ filament in Fig.\ \ref{fig:IVs}.  At room temperature, this device shows a gradual change in resistance to the compliance current (1 mA) and makes a poor memristive device, as reported previously.\cite{Sullivan2022}  As the temperature decreases, the device resistance, the threshold voltage, \vth, and the resistance off/on ratio, $R{_\textrm{off/on}}$, all increase significantly.  This increase in device resistivity is understood as not only the increase in resistance of the crystalline \NbOt\ filament, but also the suppression of the leakage current, that is, the current that flows through the amorphous matrix surrounding the \NbOt\ filament.\cite{Kwon_2024}  As the current that flows through the amorphous material decreases, the signal becomes dominated by the crystalline \NbOt\ filament, allowing the  the sharp change in resistance due to the structural IMT phase change to become apparent at about 200 K.

\begin{figure}
    \includegraphics[width=.8\linewidth]{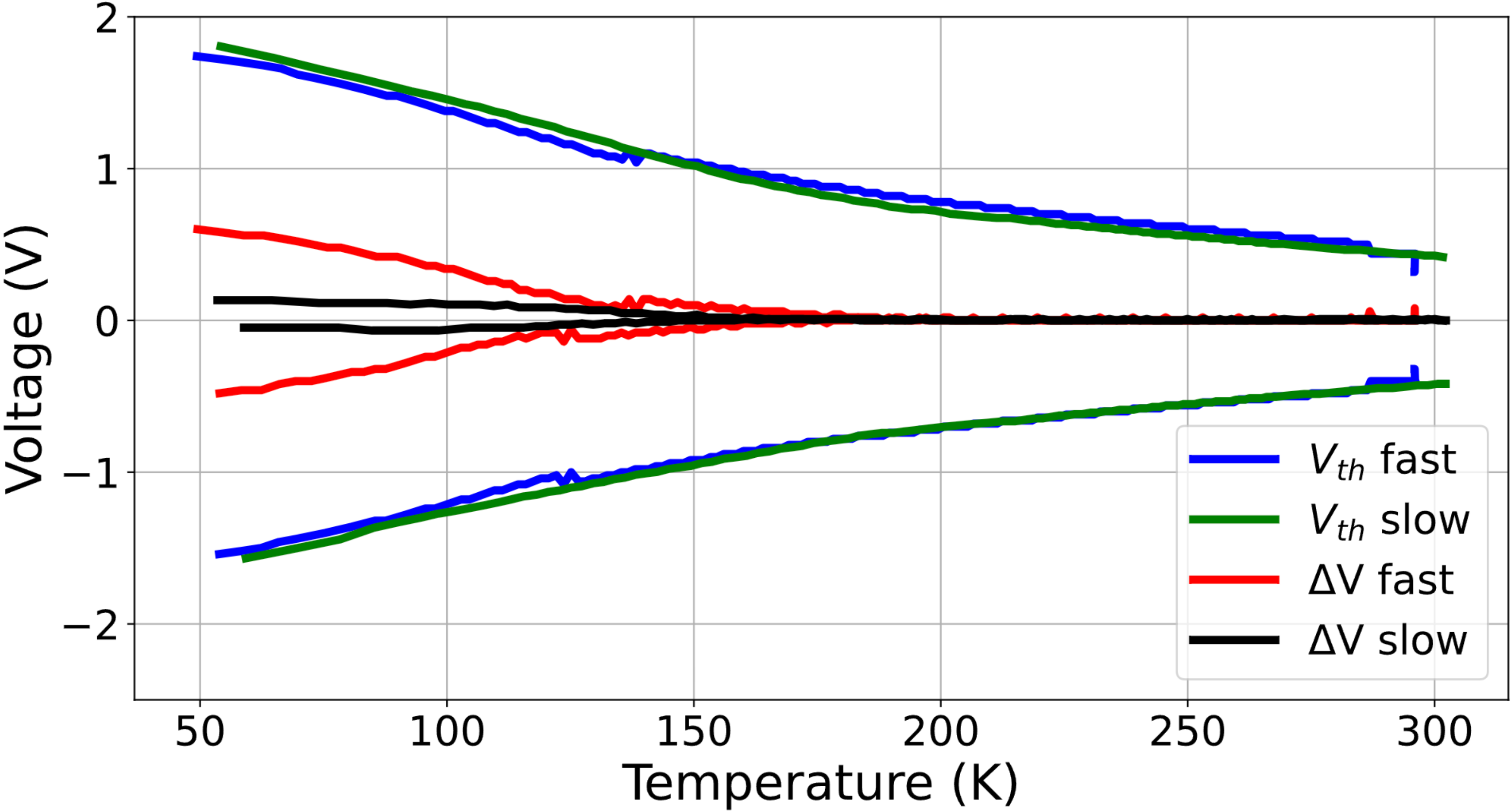} \\
    \caption{The threshold voltage \vth\ at positive and negative bias as a function of temperature.  As the temperature decreases, the \vth\ increases.  With a short integration time (200 ms) between measurements, the hold voltage \vho\ is significantly lower than the threshold voltage \vth\ and $\Delta V = |\vth - \vho|$ is large.  Turning off the drive voltage between measurements reduces $\Delta V$, but the hold voltage never reaches the threshold voltage, indicating the presence of non-volatile \NbtOf\ in the filament.}%
    \label{fig:vth}
\end{figure}

For excellent device performance, memristive devices should have a large \vth\ and a large $R{_\textrm{off/on}}$.  The change in the threshold voltage as a function of temperature is presented in Fig.\ \ref{fig:vth}.  This increase in threshold voltage has been seen in other work,\cite{Wang_2020_APL, Chen_IEEED_2021,Kwon_2024} though in the increase in \vth\ in our devices is roughly a factor of 2 larger than has been reported in the literature previously, indicating that cryogenic temperatures have improved our device performance beyond that reported previously.  This indicates that sub-stoichiometric \NbtOf\ creates devices that have a more pronounced sensitivity to temperature changes.

In addition to high threshold voltages, a large separation between the hold and threshold voltages is required for memristive device performance.  Fig.\ \ref{fig:vth} also presents our results on the hold voltage.  We expect the device to remain in the low-resistance ``on" state for voltages lower than the threshold voltage as the device remains above the IMT temperature (about 800\degC).  We find the hold voltage to be dependent on the rate of the voltage bias ramp.  A 200 ms integration time per measurement gives a $\Delta V = |\vth - \vho| \approx 0.2 V$ at or above 200 K, similar to or slightly smaller than the differences reported on previous cryogenic measurements.\cite{Wang_2020_APL, Chen_IEEED_2021,Kwon_2024}  The hold voltage stays constant as temperature decreases, leading to an increase in $\Delta V$ to nearly 0.6 V at 50 K, as shown in Fig.\ \ref{fig:vth}.  

However, longer integration times or pauses between measurements can allow the sample to cool down, which will cause the device to go through the IMT at higher hold voltages.  In Fig.\ \ref{fig:vth}, we also present data where the voltage source was turned off for 1 s between measurements, allowing the device to cool between measurements.  We show that $\Delta V$ decreases significantly, but that despite expectations, $\Delta V \neq 0$, indicating a non-volatile material embedded in series with the filament.  While \NbOt\ can mimic the volatile neuronal behavior of nerve cells, the as-grown material, \NbtOf, is attractive because it can mimic the non-volatile synaptic functions of nerve cells.\cite{Slesazeck_2015_RSC, Kumar2017_Ted_Chaos, Nath_ACSAMI_2021}  Numerical models of the IMT in \NbOt\ filaments often require the incorporation of \NbtOf\ regions to fully describe the behavior of the filaments.\cite{Liu_2016_JAP,Chen_IEEED_2021, Chen2022}  Our results support the inclusion of such non-volatile \NbtOf\ layers, as evidenced by the separation between the threshold and hold voltages, even when the device is allowed to cool down.

\subsection{Threshold Power}

\begin{figure}
    \includegraphics[width=.9\linewidth]{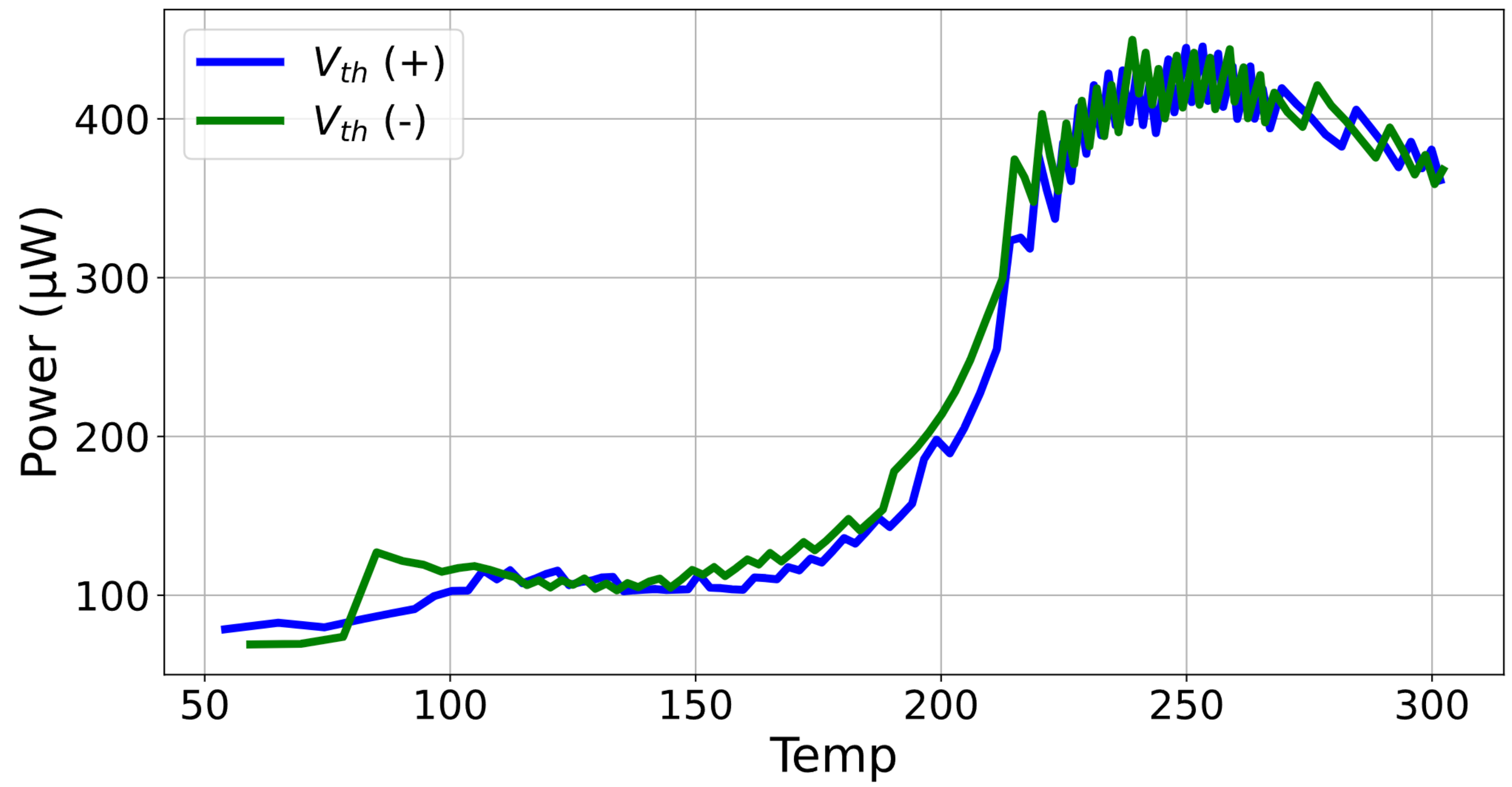} \\
    \caption{The threshold power \pth\ at positive and negative bias as a function of temperature.  As the temperature decreases, the \pth\ varies until the leakage current is suppressed, at around 200 K.  After that, the power is constant, as expected for the thermally-driven IMT in \NbOt.}%
    \label{fig:pth}
\end{figure}

The IMT in \NbOt\ is a thermally-driven transition at $\approx 800$\degC, and this high temperature is achieved by simple Joule heating in the sample.  In cross-bar filament devices, the device area can change, but the filament cross-sectional area stays constant.\cite{Sullivan2022}  This means that the power necessary to heat the material to the phase transition, \pth, also stays constant for various device sizes.\cite{Li_2018_Nano}  Because the thermal conductivity of our material stack (gold and platinum) varies by less than 10\% between 50 and 300 K, we expect the power needed to induce a thermal transition should also stay constant in our devices, even as the temperature decreases or the sample resistance increases.  Our measurements of \pth\ are presented in Fig.\ \ref{fig:pth} and show that once the leakage current is suppressed and the IMT becomes obvious in the IV curves at around 200 K, the threshold power becomes constant, as expected for the thermally driven transition.

\subsection{Resistance Measurements}
Careful study of the resistance in these devices can give a better understanding of the underlying state of \NbOt.  For low electric fields, the electrical resistivity (or conductivity) is expected to vary as:\cite{Nandi_2015_JAPD, Gibson_APL_2016, Stoever2020}
\begin{equation}
    \rho = \rho_0\, e^{E_a/k_B T},
\end{equation}
where $\rho_0$ is a constant and $E_a$ is the activation energy.  Arrhenius plots in \NbOt\ memristive devices show linear behavior and activation energies between 0.15 and 0.24 eV (when measured from 300 K to 400 K),\cite{Gibson_APL_2016, Nath_ACSAMI_2021} and measurements on thin film \NbOt\ give an Arrhenius plot that varies smoothly from $E_a =0.25$ eV at cryogenic temperatures ($\approx$ 200 K) to $E_a = 0.44$ eV near the IMT temperature.\cite{Stoever2020}


To measure the material resistance at low electric fields, researchers typically choose a voltage far from the threshold voltage, usually between at or near 0.5 V,\cite{Nath_ACSAMI_2021, Lee_JVST_2021} in order to ensure that the material is ohmic in nature.\cite{Gibson_APL_2016}  Ohmic behavior is necessary to separate the intrinsic behavior of the \NbOt\ filament from the effects of Joule heating in the sample.  We analyzed $\mathrm{d}I/\mathrm{d}V$ for our samples, and found that our sample was non-ohmic over the entire range of temperatures and voltages measured, even at the lowest measured voltages (0.03 V).  Above $\approx200$ K,  the measured resistances at the two lowest voltages (0.03 V and 0.06 V) differed by less than 1\%, indicating minimal effects due to Joule heating.  Below 200 K, the percent change in $V/I$ between the lowest measured voltages increased smoothly to more than 20\% at 80 K, indicating strong Joule heating in the sample even at very low bias voltages (0.03 V).

\begin{figure}
    \includegraphics[width=.85\linewidth]{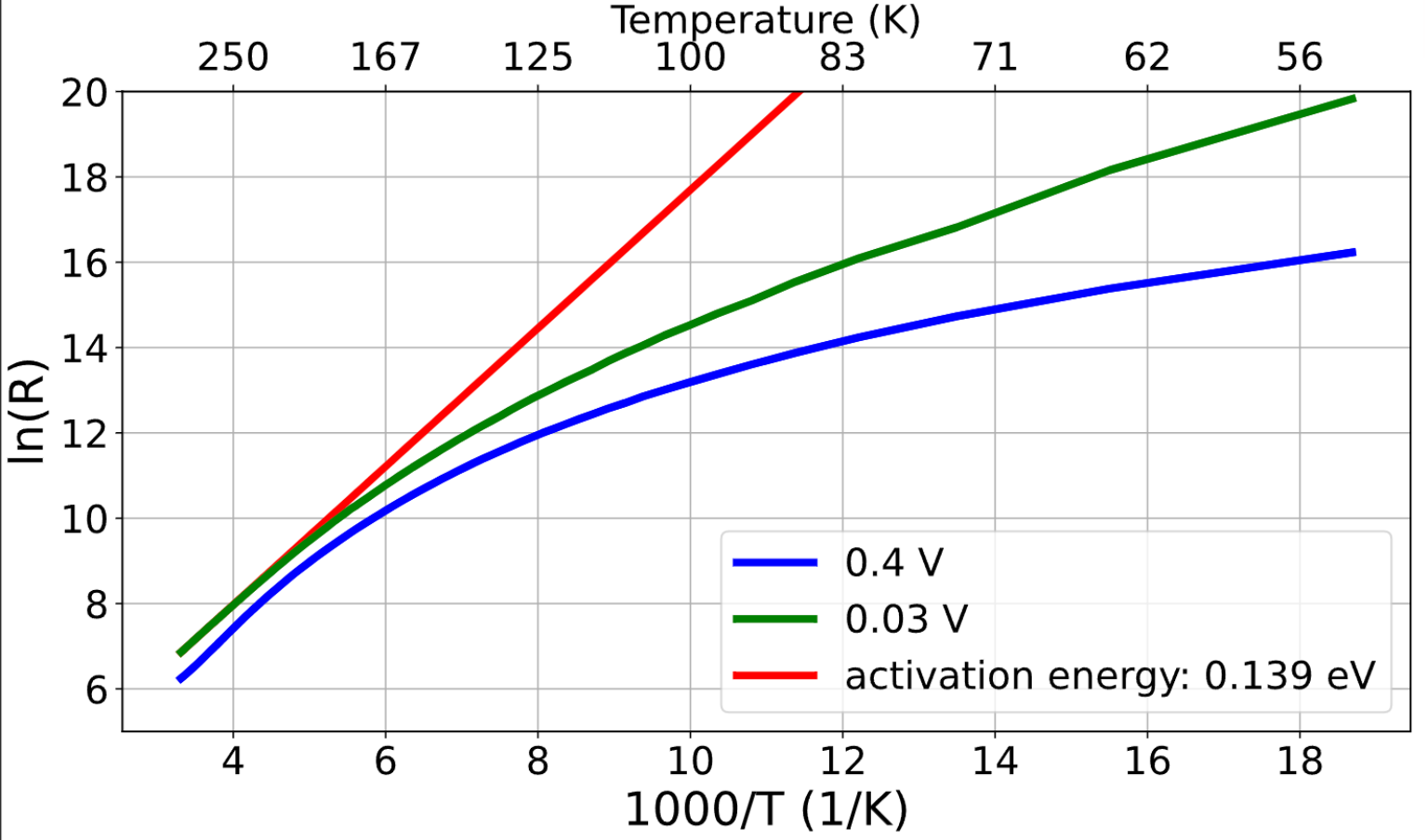} \\
    \caption{This graph plots the natural logarithm of resistance against the inverse of temperature, highlighting a smooth curve that can be compared to results from Stoever et al. (2020). Our data indicate that the temperature range of 200K - 300K is reliable for resistance measurements at low voltages, which we utilized to determine the activation energy ($E_a$) of the \NbOt.}%
    \label{fig:lnR}
\end{figure}

Our measurements of the device resistance are presented in an Arrhenius plot in Fig.\ \ref{fig:lnR}.  We presented $V/I$ at the lowest measured resistance, 0.03 V, as well as a conventional choice, 0.4 V.  Above 200 K, the Arrhenius curve is linear for 0.03 V, and we find an activation energy of $E_a = 0.14$ eV.  This activation energy is lower than the $E_a \approx 0.2$ eV reported for \NbOt\ devices,\cite{Nath_ACSAMI_2021, Gibson_APL_2016} and significantly lower than$E_a \sim 0.3-0.4$ eV reported for thin-film \NbOt.\cite{Stoever2020}  Below 200 K in Fig.\ \ref{fig:lnR}, we see a smoothly varying curve for 0.03 V, similar to results in thin-film \NbOt,\cite{Stoever2020} and due to the significant Joule heating, a line fit is not possible. 

In order to compare our devices to others reported in the literature, we have also included the measured $V/I$ at 0.4 V in Fig. \ref{fig:lnR}.  At 0.4 V, the device has a lower resistance compared to the measurements at 0.03 V, which we expect, as the higher temperature at 0.4 V due to Joule heating leads to higher device temperatures.  Above $\approx 225$ K, the slope at 0.4 V matches the slope at 0.03 V, meaning the activation energy is similar at $T \gtrsim 225$ K.

The results in our filament devices are correlated: The lower activation energy in the device leads directly to both the lower resistance and the poor switching behavior when compared with other filament devices at room temperature. As a result, we can conclude that as-grown sub-stoichiometric  \NbtOf\ is not a good candidate for \NbOt\ devices via electroformation.  The low activation energy in this material means that the resistance in these devices must be increased before it can be useful for memristors \textemdash this can be accomplished by decreasing the temperature, or by decreasing the device size.\cite{Sullivan2022}

\section{Conclusion}

We have studied filamentary devices electroformed from as-deposited sub-stoichiometric \NbtOf{}.  At room temperature, these devices are poor candidates for memristive applications, with low threshold voltages and with an $R_\mathrm{off/on}$ ratio near zero.\cite{Sullivan2022}  

Cryogenic measurements of these same devices show a marked increase in resistance due to the suppression of leakage currents.  Below 200 K, the filaments display a large increase in the threshold voltage driven by the increase in the device resistance at lower temperatures.  Nevertheless, the power required to heat the sample to the IMT temperature stays constant, as expected for this thermally-driven transition.  The off/on resistance ratio increases to values between 10 and 50, and the hold voltage is significantly lower than the threshold voltage for rapid voltage sweeps.  Thus, at cryogenic temperatures, the device performance improves significantly.

There is a wide variety of \NbOt\ and \NbtOf\ films currently being used by researchers in efforts to optimize the niobium oxide's performance for memristive devices.  Some of the widely varying results come from the lack of clarity in growth conditions and lack of clarity in the underlying niobium oxide films used for devices.  However, our results indicate that even nominally ``bad" films can be used for memristive devices if steps are taken to reduce the leakage current and increase the device resistance: through percolation paths and incomplete annealing,\cite{Lee_JVST_2021} through annealing and nanoscale device sizes,\cite{Sullivan2022} or through operation at cryogenic temperatures.

\begin{acknowledgments}
This research was supported by the National Science Foundation grants nos.\ DMR-2103197, DMR-2103185 and the Air Force Research Laboratory grant FA8750-21-1-1019. 

This work was performed in part at the Cornell NanoScale Facility, a member of the National Nanotechnology Coordinated Infrastructure (NNCI), which is supported by the National Science Foundation (Grant NNCI-2025233) and made use of the Cornell Center for Materials Research Shared Facilities, which are supported through the NSF MRSEC program (DMR-1719875). The niobium oxide films were grown at the Albany Nanotech Complex via NY CREATES and Dr.\ Sandra Schujman at NY CREATES assisted with x-ray measurements.

\end{acknowledgments}

\vspace{10 pt}

\noindent \large{\textbf{{Data Availability}}
\normalsize
The data that support the findings of this study are available from the corresponding author upon reasonable request.

\vspace{10 pt}

\noindent \large{\textbf{{Conflicts of Interest}}
\normalsize
The authors have no conflicts to disclose.


%
%

%

\bibliography{NbO2_cryogenic}

\end{document}